\documentclass[aip,jap,amsmath,amssymb,preprint,floatfix]{revtex4}
\usepackage{graphicx}% Include figure files
\usepackage{dcolumn}% Align table columns on decimal point
\usepackage{bm}% bold math
\usepackage[utf8]{inputenc}
\usepackage[T1]{fontenc}
\usepackage{mathptmx}
\usepackage{etoolbox}
\usepackage{xcolor}
\usepackage{siunitx}
\usepackage[version=4]{mhchem}
\usepackage{mathrsfs}
\usepackage[english]{babel}
\usepackage{mathtools}

\makeatletter
\def\@email#1#2{%
 \endgroup
 \patchcmd{\titleblock@produce}
  {\frontmatter@RRAPformat}
  {\frontmatter@RRAPformat{\produce@RRAP{*#1\href{mailto:#2}{#2}}}\frontmatter@RRAPformat}
  {}{}
}%
\makeatother

\graphicspath{figures/}

\newcommand{\ex}[1]{\exp\paren{#1}}
\newcommand{\paren}[1]{\left(#1\right)}

\begin{document}

%\preprint{}
\title[Memristor Compact Model]{Memristor Compact Model with Oxygen-Vacancy Concentration as State Variable}

\author{Andre Zeumault$^*$}
\email{azeumault@utk.edu}
\altaffiliation[Also at ]{Department of Materials Science and Engineering, University of Tennessee, Knoxville, TN, USA, 37996}
\author{Shamiul Alam}
\author{Md Omar Faruk}
\author{Ahmedullah Aziz}
\affiliation{Department of Electrical Engineering and Computer Science, The University of Tennessee, Knoxville, TN, USA, 37996}

\date{\today}

\begin{abstract}
We present a unique compact model for oxide memristors, based upon the concentration of oxygen vacancies as state variables. In this model, the increase (decrease) in oxygen vacancy concentration is similar in effect to the reduction (expansion) of the tunnel gap used as a state variable in existing compact models, providing a mechanism for the electronic current to increase (decrease) based upon the polarity of the applied voltage. Rate equations defining the dynamics of state variables are obtained from simplifications of a recent manuscript in which electronic processes (i.e., electron capture/emission) were combined with atomic processes (i.e., Frenkel pair generation/recombination, diffusion) stemming from the thermochemical model of dielectric breakdown. Central to the proposed model is the effect of the electron occupancy of oxygen vacancy traps on resistive switching dynamics. The electronic current is calculated considering Ohmic, band-to-band, and bound-to-band contributions. The model includes uniform self-heating with Joule-heating and conductive loss terms. The model is calibrated using experimental current-voltage characteristics for \ce{HfO2} memristors with different electrode materials. Though a general model is presented, a delta-shaped density of states profile for oxygen vacancies is found capable of accurately representing experimental data while providing a minimal description of bound to band transitions. The model is implemented in Verilog-A and tested using read/write operations in a 4x4 1T1R nonvolatile memory array to evaluate its ability to perform circuit simulations of practical interest. A particular benefit is that the model does not make strong assumptions regarding filament geometry of which scant experimental-evidence exists to support. 
\end{abstract}

\maketitle

\section{Introduction}

Several physics-inspired compact models have been proposed for circuit-simulation of oxide memristors--key elements for emerging applications in non-volatile memories, unconventional computing, and biologically-inspired computing alike~\cite{raj_memristor_2021,du_low-power_2021,sun_hybrid_2010,thomas_memristor-based_2013,lu_biological_2020}. These are largely based on assumptions regarding the geometry of a ``conductive filament'', described either in terms of a one-dimensional tunnel gap~\cite{chen_synapse_2013,zhang_memristive_2017, vourkas_spice_2015,guan_SPICE_2012,chen_compact_2015} or the size/shape of the filament approximated as cylindrical~\cite{bocquet_robust_2014} or rectangular~\cite{biolek_spice_2009,kvatinsky_team_2013}. Electronic conduction processes--derived from changes in filament geometry--are usually modeled with drift/diffusion~\cite{gao_oxide-based_2008,larentis_resistive_2012,gilmer_asymmetry_2012,vandelli_modeling_2011,nardi_resistive_2012}, hopping\cite{huang_physics-based_2013}, trap-assisted and band-to-band tunneling\cite{guan_SPICE_2012}, Poole-Frenkel emission, site-percolation\cite{sune_new_2001,sune_analytical_2009,tous_compact_2010,long_compact_2013}, or interfacial redox reactions \cite{bocquet_robust_2014}. Additionally, conduction models sometimes incorporate simple thermal models (e.g., uniform Joule heating) since local temperature is purported to have a significant impact on electrical characteristics, particularly retention~\cite{kumar_Conduction_2016} and multibit operation~\cite{alexandrov_current-controlled_2011}. 

The scope of existing tunnel-gap models can be understood simply by taking any or all of the nominal conduction processes present in metal-insulator-metal devices (e.g., direct and Fowler-Nordheim tunneling, thermionic emission, Poole-Frenkel emission, Ohmic conduction, ionic conduction and space-charge limited conduction~\cite{sze_physics_2006}) and replacing the insulator thickness parameter with a dynamic variable (i.e., the tunnel gap) ranging from zero to a few nanometers~\cite{guan_SPICE_2012}. The dynamics of the tunnel gap are based on the diffusion kinetics of oxygen ions presumed to be rate-limiting--approximated in one-dimension along the oxide thickness using the Mott-Gurney drift velocity~\cite{yu_Phenomenological_2010}. A positive voltage drives ions towards the top electrode, reducing the tunnel gap; a negative voltage drives ions towards the bottom electrode, increasing the tunnel gap. Since this implicitly assumes Frenkel pairs recombine (generate) in the presence (absence) of oxygen ions, the thermal barriers for recombination/generation must be smaller than that of oxygen ion diffusion wherever the model is applied so as not to become rate-limiting. The zero-field thermal barrier for Frenkel pair generation can be \SI{1.4}{\electronvolt} (at grain boundary sites~\cite{aldana_Resistive_2020}), \SI{2.2}{\electronvolt} (at gettering electrode surfaces~\cite{xu_Kinetic_2020}) and several electronvolts in the bulk (\SI{3.6}{\electronvolt}~\cite{aldana_Resistive_2020}, \SI{6.8}{\electronvolt}~\cite{xu_Kinetic_2020}) all of which are much larger than that of oxygen ion diffusion ($\approx$ \SI{0.7}{\electronvolt}~\cite{kumar_Conduction_2016}). Consequently, tunnel gap models are adequate for describing reset or set operations~\cite{jiang_Verilog_2014,guan_SPICE_2012,yu_Phenomenological_2010} particularly when the top electrode can be regarded as an oxygen reservoir with permeable interface and the electric field is large enough (tunnel gap small enough) that the thermal barrier for Frenkel pair generation is reduced and no longer rate-limiting. A more complex or entirely new framework would be required to incorporate a physical description of forming self-consistently within this model~\cite{guan_Switching_2012}. Given that oxide memristors are flux driven~\cite{chua_Memristorthe_1971}, a description which takes into account the entire electrical history of the device (forming, reset and set) is critical from a self-consistent physical modeling perspective.

It is important to note that all compact models are phenomenological, involving many simplifications to enable circuit simulation at various levels of complexity and computational efficiency. Ultimately, the accuracy and speed of the device compact model is selected based on the needs of the circuit designer rather than the physical accuracy alone. For example, simple piecewise linear models may be desired when the focus is on developing larger neuromorphic systems~\cite{amer_practical_2017} whereas predictive physics-based models may be desired for testing smaller memory array architectures~\cite{pickett_switching_2009}. Ideally, memristor compact models would stem from a common physical framework in which complexity can be selectively refined at the circuit level, similar to the different levels of the well-known SPICE models for conventional bulk MOSFETs (square law, bulk charge, etc.).

In this work, a compact model for oxide memristors is presented that is based on the concentration of oxygen vacancies and their electron occupancy as opposed to one-dimensional filament geometry. The shift from a tunnel gap to a concentration state variable may be more physically appropriate, given that the formation of a one-dimensional defect chain is statistically unlikely considering other competing factors such as vacancy diffusion, thermophoresis~\cite{kumar_Conduction_2016} and the increased thermodynamic stability of defect clusters~\cite{bradley_ElectronInjectionAssisted_2015} expected to lead to additional conduction pathways. These, along with the presence of grain-boundaries and other imperfections which may reduce the zero-field formation enthalpy of oxygen vacancies suggest a variety of filament shapes with varying connectivity as seen in comprehensive modeling approaches~\cite{aldana_3D_2017,guan_Switching_2012,zeumault_tcad_2021} and experimental observations~\cite{li_Direct_2017}. 

In our model, the oxygen vacancy concentration is split into two state variables to allow the electron-occupancy of the vacancies to vary between occupied and unoccupied depending on electron capture and emission processes. This is based on recent KMC modeling approaches~\cite{zeumault_tcad_2021} and reflects ab-initio calculations showing increased thermodynamic stability of negatively-charged oxygen vacancies under conditions of electron-injection~\cite{bradley_ElectronInjectionAssisted_2015} as well as experimental observations of a negative-space charge associated with suspected filament regions~\cite{li_Direct_2017}. This has immediate implications on retention, since a larger thermal emission barrier would result in a more stable low resistance state and therefore, longer retention time as we show by comparison to experimental data of memristors having different electrode materials. The compact model is implemented in Verilog-A, using simplifying assumptions made to our previous model \cite{zeumault_tcad_2021}. In particular, a key approximation is that the field-dependence of the microstate transitions is approximated using the average electric field as opposed to the local electric field. Since electron capture and emission processes are taken into account, the model utilizes additional parameters associated with the electrodes (e.g., work function, effective mass) and those of the oxygen vacancy defects (e.g., trap energy level/ionization energy, capture cross-section, and thermal capture barrier) that are intimately connected to the resistive switching dynamics. The model also considers temperature effects and, most importantly, does not assume a particular filament shape, gap size between filament tip or any other restrictive geometrical constraints which are not firmly supported by experimental data. Each of these features make this compact model attractive due to its predictive modeling capability and pairing ability with ongoing experimental investigations to identify and observe filament geometries~\cite{kumar_Conduction_2016,li_Direct_2017}. Being generally based upon a concentration of defects as opposed to a particular filament geometry, the model is open to different interpretations of conduction (e.g., drift/diffusion, trap-assisted-tunneling or percolation) as we show. 

Lastly, the treatment of oxygen vacancies as an electron trap allows the material-specific density of states to be specified to provide increased accuracy/reduced efficiency (continuous energy dependence) or decreased accuracy/increased efficiency (discrete trap level) based on the particular needs of the circuit designer. The density of states is more intimately connected to materials chemistry than is a tunneling gap, allowing greater flexibility when comparing model predictions to experimental data (e.g., defect spectroscopy~\cite{thamankar_Single_2016}). For circuit-simulation, a delta density of states profile is least mathematically complex to implement, requiring no numerical integration at each time step. Thus we extensively evaluate the compact model (implemented in Verilog-A) based on a delta profile density of states. The model is validated using independent experimental data-sets having different electrode materials and tested using a simple 4x4 1T1R array representative of a small nonvolatile memory array to evaluate the model's capability to perform basic circuit simulations. 

\section{Compact Model Framework}

\subsection{General Assumptions and Notation}

\begin{table}
\begin{center}\begin{tabular}{lll}
\hline
Parameter & Description & Value and Unit\\
\hline
$q$ & Elementary charge & \SI{1.602176e-19}{\coulomb} \\
$k$ & Boltzmann constant & \SI{1.380649e-23}{\joule\per\kelvin} \\
$\hbar$ & Reduced Planck constant & \SI{1.054571e-34}{\joule\second} \\
$T_0$ & Ambient temperature & \SI{300}{\kelvin} \\
$\epsilon_r$ & Relative permittivity & \SI{30}{} \\
$R_{0}$ & Attempt-to-escape frequency & \SI{10}{\tera\hertz} \\
$E_{a,gen}^0$ & Activation energy for Frenkel pair generation & \SI{7.3}{\electronvolt} \\
$E_{a,rec}^0$ & Activation energy for Frenkel pair recombination & \SI{0.15}{\electronvolt} \\
$E_{ion}$ & Ionization energy of oxygen vacancy defect & \SI{2.0}{\electronvolt} - \SI{3.0}{\electronvolt}\\
$\sigma_0$ & Capture cross-section of oxygen vacancies & \SI{1e-14}{\centi\meter\squared}\\
$\Phi^{TE}$ & TE work function (\ce{Ti}) & \SI{4.33}{\electronvolt} \\
$\Phi^{BE}$ & BE work function (\ce{TiN}) & \SI{4.50}{\electronvolt} \\
$m_e^{TE}$ & TE electron effective mass (\ce{Ti}) & \SI{3.2}{} \\
$m_e^{BE}$ & BE electron effective mass (\ce{TiN}) & \SI{2.0}{} \\
$n_e^{TE}$ & TE electron concentration (\ce{Ti}) & \SI{5.67e22}{\per\centi\meter\cubed}\\
$n_e^{BE}$ & BE electron concentration (\ce{TiN}) & \SI{2.88e22}{\per\centi\meter\cubed}\\
$m_e^{ox}$ & Oxide electron effective mass (\ce{HfO2}) & \SI{0.1}{} \\
$\vec{p}$ & Oxide molecular dipole moment (\ce{HfO2}) & \SI{15e-10}{\coulomb\meter}\\
$E_g$ & Oxide bandgap energy (\ce{HfO2}) & \SI{5.9}{\electronvolt}\\
$\chi$ & Oxide electron affinity (\ce{HfO2}) & \SI{2}{\electronvolt} \\
$\mu_{eff}$ & Effective band mobility (\ce{HfO2}) & \SI{1}{\centi\meter\squared\per\volt\per\second}\\
$c_{p,ox}$ & Oxide specific heat capacity (\ce{HfO2}) & \SI{120}{\joule\per\kilogram\per\kelvin}\\
$C_{add.}$ & Addenda heat capacity (Device) & \SI{1.45e-9}{\joule\per\kelvin}\\
$\rho_{ox}$ & Oxide mass density (\ce{HfO2}) & \SI{9800}{\kilogram\per\meter\cubed}\\
$\kappa_{ox}$ & Oxide thermal conductivity (\ce{HfO2}) & \SI{1}{\watt\per\meter\per\kelvin}\\
$N_{sites}$ & Site density (\ce{HfO2}) & \SI{4.38e19}{\per\centi\meter\cubed}\\
$y_t$ & Trap center (\ce{HfO2}) & $0.5t_{ox}$\\
\hline
\end{tabular}\end{center}
\caption{Summary of nominal parameters used in this work unless otherwise stated.}
\label{tbl:parameters}
\end{table}

\begin{figure*}
\includegraphics[width=\textwidth]{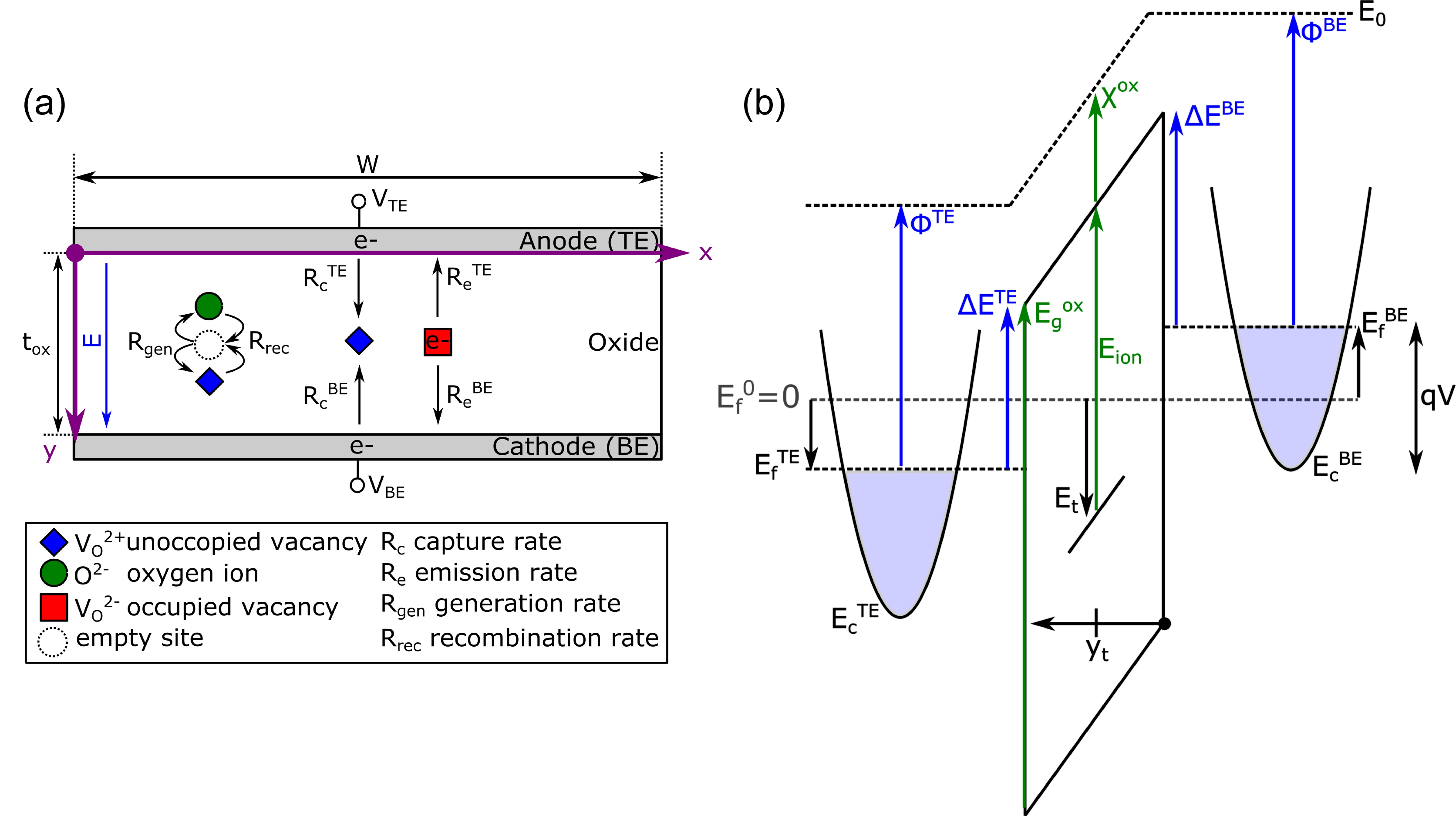}
\caption{(a) Device structure, showing the physical processes and memristor components modeled herein. (b) Band diagram showing definitions of energy parameters. (not to scale)}
\label{fig:description}
\end{figure*}

In Table~\ref{tbl:parameters}, we define the notation for symbols used throughout the model as well as nominal material parameters for \ce{HfO2} which we evaluate herein as the oxide material. In Figure~\ref{fig:description}, the device structure and definitions of energy levels are depicted. Although the model is presented using parameters for \ce{HfO2}, it is generally applicable to transition metal oxides of interest for memristors since it is based upon a general thermochemical model of dielectric breakdown~\cite{mcpherson_thermochemical_2003,mcpherson_underlying_1998}, common to many insulating oxides and/or polar solids in general.

Below, we summarize the main assumptions used in deriving the compact model:
\begin{itemize}
\item The oxide is single-phase (i.e., no polymorphism) with material properties of the single-crystalline bulk (e.g., effective mass, permittivity, bandgap, electron affinity). In practice, this implies the material properties are homogeneous and is not necessarily tied to the bulk crystalline or amorphous state.
\item The local electric field can be approximated with the spatially-averaged field.
\item Oxygen vacancies behave as electron traps (positive when empty $V_O^{2+}$, negative when occupied $V_O^{2-}$, accommodating 4 electrons~\cite{bradley_ElectronInjectionAssisted_2015}), defined in terms of a continuous electronic density of states to specify the ionization energies/defect levels participating in electron capture/emission.
\item The interface tunneling transmission probabilities for the top and bottom electrodes are approximately equal.
\item Diffusion of ionic point defects is not considered.
\item Trap to trap coupling is not considered.
\end{itemize}

\subsection{State Variable Dynamics}

The state of the oxide is represented using three species: unoccupied oxygen vacancies $N_{V_O^{2+}}$, occupied oxygen vacancies $N_{V_O^{2-}}$, and empty states $N_{0}$; each of these are taken to be volume-averaged quantities (i.e., concentrations). These are depicted in Figure~\ref{fig:description}(a). At any point in time, the sum of these is constant and equal to the total density of accessible sites, which is nominally equal to the atomic density of \ce{HfO2} ($N_{sites} \approx \SI{2.77e22}{\per\centi\meter\cubed}$). 
\begin{equation}
N_0 + N_{V_O^{2+}} + N_{V_O^{2-}} = N_{sites}
\label{eqn:sum}
\end{equation}

We note that, in general, $N_{sites}$ can be regarded as a parameter which establishes an upper bound for the state variables $N_{V_O^{2+}}$, $N_{V_O^{2-}}$, and $N_0$. When $N_{sites}$ is equal to the atomic density, the intrinsic state of the memristor is reachable in the absence of external current-compliance. As will be shown, reducing $N_{sites}$ below the atomic density has a similar effect as forcing a current-compliance. A value of \SI{4.38e19}{\per\centi\meter\cubed} was used throughout this work based on agreement to experimental data.

From~\ref{eqn:sum}, since $N_{sites}$ is a constant, the net rate of change in each species must sum to zero.
\begin{equation}
\frac{dN_0}{dt} + \frac{dN_{V_O^{2+}}}{dt} + \frac{dN_{V_O^{2-}}}{dt} = 0
\label{eqn:conservation}
\end{equation}

These rates can be expressed in terms of the following physical transitions: Frenkel-pair generation $R_{gen}$ and recombination $R_{rec}$, electron capture $R_c$ and emission $R_e$ in which capture/emission of electrons can take place between either the top electrode (TE) or bottom electrode (BE). 

The rate of change in electron occupancy of oxygen vacancy traps, due to electron capture and emission is given by Equation~\ref{eqn:rate-traps}.
\begin{align}
\frac{df^{t}}{dt} &= \paren{1-f^{t}}\underbrace{\paren{f^{BE}R_c^{BE}+f^{TE}R_c^{TE}}}_{R_c}-f^{t}\underbrace{\paren{(1-f^{BE})R_e^{BE}+(1-f^{TE})R_e^{TE}}}_{R_e} \nonumber \\
&= \paren{1-f^{t}}R_c - f_tR_e
\label{eqn:rate-traps}
\end{align}

In terms of $f^t$ and the density of states $g(E_t)$, the occupied, unoccupied and total vacancy concentration can be calculated using Fermi-Dirac statistics.
\begin{equation}
N_{V_O^{2-}} = \int f^t g(E_t) dE_t
\label{eqn:occupied_vacancies}
\end{equation}
\begin{equation}
N_{V_O^{2+}} = \int (1-f^t) g(E_t) dE_t
\label{eqn:unoccupied_vacancies}
\end{equation}
\begin{equation}
N_{V_O} = N_{V_O^{2+}} + N_{V_O^{2-}} = \int g(E_t) dE_t
\label{eqn:total_vacancies}
\end{equation}

Trap-to-trap coupling is neglected, which would otherwise give rise to a separate Fermi level for trap states $E_f^{t}$ and corresponding Fermi function $f^{t}$ for use in calculating trap concentrations. Instead, the number density of traps is used to compute the electron occupancy probability.
\begin{align}
f^t &= \frac{N_{V_O^{2-}}}{N_{V_O}} \\
1-f^t &= \frac{N_{V_O^{2+}}}{N_{V_O}}
\end{align}

Using these definitions, integration of Equation~\ref{eqn:rate-traps} over the density of trap states $g(E_t)$ provides the rate equation for the occupied vacancy concentration $N_{V_O^{2-}}$,
\begin{align}
\frac{dN_{V_O^{2-}}}{dt} &= \frac{1}{N_{V_O}}\int g(E_t)\paren{N_{V_O^{2+}}R_c-N_{V_O^{2-}}R_e}dE_t \nonumber \\
&= \int \hat{g}(E_t)\paren{N_{V_O^{2+}}R_c-N_{V_O^{2-}}R_e}dE_t 
\label{eqn:master}
\end{align}
where $\hat{g}(E_t)$ is the normalized density of states profile for the oxygen vacancy levels.
\begin{equation}
\hat{g}(E_t) = \frac{g(E_t)}{\int g(E_t) dE_t} = \frac{g(E_t)}{N_{V_O}}
\end{equation}

Equation~\ref{eqn:master} simply states that unoccupied vacancy states $N_{V_O^{2+}}$ are filled by either electrode (TE/BE) and occupied states $N_{V_O^{2-}}$ are emptied to either electrode (TE/BE). The energy integral allows for the density of states to take on any energy dependence (e.g., delta, uniform, gaussian). 

To account for defect creation, which initially gives rise to the formation of Frenkel pairs consisting of oxygen ions and unoccupied oxygen vacancies (i.e., $O^{2-}$, $V_O^{2+}$), atomic processes are incorporated by assuming that the empty states in our model $N_0$ are increased (decreased) due to recombination (generation) of Frenkel pairs. 
\begin{equation}
    \frac{dN_0}{dt} = R_{rec}N_{V_O^{2+}} - R_{gen}N_{0}
    \label{eqn:empty}
\end{equation}

Using Equation~\ref{eqn:conservation}, along with Equations~\ref{eqn:master} and \ref{eqn:empty}, the remaining rate equation for the unoccupied vacancies is given by Equation~\ref{eqn:unoccupied}.
\begin{align}
    \frac{dN_{V_O^{2+}}}{dt} &= R_{gen}N_{0} - R_{rec}N_{V_O^{2+}} - \frac{dN_{V_O^{2-}}}{dt} \nonumber \\
    &= R_{gen}N_{0} - R_{rec}N_{V_O^{2+}} - \int \hat{g}(E_t)\paren{N_{V_O^{2+}}R_c-N_{V_O^{2-}}R_e}dE_t
    \label{eqn:unoccupied}
\end{align} 

Together, Equations~\ref{eqn:master}, \ref{eqn:empty} and \ref{eqn:unoccupied} define a dynamic relationship between \textit{atomic} and \textit{electronic} processes, which we have evaluated numerically in a recent manuscript in two dimensions, taking into account diffusion, Frenkel pair generation and recombination, electron capture and emission~\cite{zeumault_tcad_2021}. For simplicity, the compact model developed here does not consider diffusion; diffusion of charged point-defects would produce a spatially and time-dependent electric field which would complicate compact model development. 

\subsection{Transition Rates}

The rates $R_{gen}$ and $R_{rec}$ are atomic processes describing the creation and removal of Frenkel pairs of oxygen ions and vacancies (i.e., $O^{2-}$ and $V_O^{2+}$) whereas the rates $R_c$ and $R_e$ are electronic processes describing the capture and emission of electrons by oxygen vacancies, to either electrode. 

The rates of Frenkel-pair generation/recombination can be described according to classical transition state theory,  commonly applied in kinetic Monte Carlo simulations of oxide memristors.
\begin{align}
R_{gen} &= R_0\ex{-\frac{E_{a,gen}}{kT}}\\
R_{rec} &= R_0\ex{-\frac{E_{a,rec}}{kT}}
\label{eqn:gen_rec}
\end{align}
The thermal barriers for these transitions are electric field dependent (the local field can promote charge separation or recombination), which we model as a symmetric raising/lowering of the thermal barrier with field $\mathscr{E}$.
\begin{align}
    E_{a,gen} &= E^0_{a,gen} - \mathscr{E}|\vec{p}|\paren{\frac{\epsilon_r + 2}{3}} \\
    E_{a,rec} &= E^0_{a,rec} + \mathscr{E}|\vec{p}|\paren{\frac{\epsilon_r + 2}{3}}
    \label{eqn:thermochemical}
\end{align}
These equations are implemented such that the maximum rate cannot exceed the exponential prefactor $R_0$, occuring for high positive/negative voltages, otherwise leading to negative thermal barriers. 

The capture rates $R_c^{TE/BE}$ can be described using a classical or quantum approach. The latter uses Fermi's Golden Rule for the rate of interaction $R_{inter.}$ between the bound state (within the oxide) and the band state (within the electrode). For simplicity, the interaction potential can be confined to a box-shaped volume $V_t$ in which the interaction potential energy is set to the defect level referenced to the conduction band $E_t = E_c - E_{ion}$. For 2D or 3D modeling, in which spatially-dependent features are integral, a quantum approach may be desired for its improved accuracy. A complete derivation of both is provided in the supplementary material of our previous work~\cite{zeumault_tcad_2021} and elsewhere~\cite{jimenez-molinos_direct_2002,palma_quantum_1997} as implemented in standard device simulators~\cite{synopsys_TCAD_2019}. 

Equation~\ref{eqn:capturerate-quantum} shows the resulting quantum model for an elastic transition.
\begin{align}
    R_c^{TE/BE} &= P_{tun.}R_{inter.} \nonumber\\ 
    &= \ex{-\frac{y_t}{y_0^{TE/BE}}}\frac{\sqrt{8}}{\hbar^4\pi}(m^{TE/BE}_{e})^{3/2}V_tE_{ion}^2\sqrt{E_t-E_c^{TE/BE}}f(E_t)^{TE/BE}
    \label{eqn:capturerate-quantum}
\end{align}
An additional factor is included to account for the tunneling probability $P_{tun.}$, using the Wentzel-Kramers-Brillouin (WKB) approximation for a triangular barrier of uniform field~\cite{sze_physics_2006}. Here, $f(E)^{TE/BE}$ is the Fermi-Dirac function, defined separately for each electrode; $y_t$ is the position of the trap relative to the electrode from which an electron is captured; $m_e^{TE/BE}$ is the electron effective mass of the electrode; $V_t$ is the trap volume; $y_0^{TE/BE}$ is a WKB tunneling parameter~\cite{sze_physics_2006}. $V_t$ and $y_0$ can be expressed using a box-shaped potential for the trap state and a triangular barrier (uniform field) approximation for the conduction band as~\cite{jimenez-molinos_direct_2002,palma_quantum_1997}
\begin{align}
V_t &= \paren{\frac{\hbar}{\sqrt{2m_e^{ox}E_{ion}}}}^3\\
y_0 &= \frac{3}{4}\frac{\hbar}{\sqrt{2m_e^{TE/BE}\Delta E^{TE/BE}}}
\end{align}
where $\Delta E^{TE/BE} = \Phi^{TE/BE}-\chi$ is the conduction band offset at each electrode interface, $\Phi^{TE/BE}$ is the electrode work function and $\chi$ is the electron affinity of the oxide; $y_0$ has two values, one for each electrode.

Equation~\ref{eqn:capturerate-quantum} assumes that the interaction between bound states and band states is elastic, and that the electrodes can be approximated using a bulk parabolic band dispersion. 

Here, for simplicity, we implement a classical model for the rate of interaction in terms of a capture cross-section, thermal velocity, carrier concentration and thermal barrier. The classical model has the same phenomenological aspects of the quantum model, though it differs quantitatively regarding the dependence on the trap level and applied voltage. 

Equation~\ref{eqn:capturerate-classical} shows the resulting classical model. 
\begin{equation}
R_c^{TE/BE} = R_{c,0}^{TE/BE} \ex{-\frac{y_t}{y_0^{TE/BE}}}\ex{-\frac{E_{a,t}}{k_BT}}
\begin{dcases}
\ex{\frac{\mp q\mathscr{E}y_t-E_{tf}}{k_BT}} & E_{tf}^{TE/BE} > 0\\
\ex{\frac{\mp q\mathscr{E}y_t}{k_BT}} & E_{tf}^{TE/BE} < 0\\
\end{dcases}
\label{eqn:capturerate-classical}
\end{equation}
where $R_{c,0}^{TE/BE}$ and $E_{tf}^{TE/BE}$ are defined as,
\begin{align}
R_{c,0}^{TE/BE} &= \sigma_0 v_{th}^{TE/BE} n_{e}^{TE/BE} \\
E_{tf}^{TE/BE} &= E_t - E_f^{TE/BE} \\
v_{th}^{TE/BE} &= \sqrt{\frac{3kT}{m^{TE/BE}_e}}
\end{align}

Regarding the $\mp$ in front of the field-dependent terms in Equation~\ref{eqn:capturerate-classical}, the $-$ is used for the top electrode (TE) and the $+$ is used for the bottom electrode (BE). In either model, the emission rate to either electrode is calculated from the capture rates by the detailed balance criterion.
\begin{equation}
    R_e^{TE/BE} = \ex{-\frac{E^{TE/BE}_{f}-E_t}{kT}}R_c^{TE/BE}
\end{equation}
For the classical model, this leads to the following,
\begin{align}
R_e^{TE/BE} &= R_{c,0}^{TE/BE} \ex{-\frac{y_t}{y_0^{TE/BE}}}\ex{-\frac{E_{a,t}}{k_BT}}
\begin{dcases}
\ex{\frac{\pm q\mathscr{E}y_t}{k_BT}} & E_{tf} > 0\\
\ex{\frac{\pm q\mathscr{E}y_t+E_{tf}}{k_BT}} & E_{tf}^{TE/BE} < 0\\
\end{dcases}
\label{eqn:emissionrate-classical}
\end{align}
Regarding the $\pm$ in front of the field-dependent terms in Equation~\ref{eqn:emissionrate-classical}, the $+$ is used for the top electrode (TE) and the $-$ is used for the bottom electrode (BE). 

The voltage dependence of these expressions enters via a symmetric raising/lowering of the Quasi Fermi energies from the equilibrium Fermi level (taken as zero, $E_f^{TE,0} = E_f^{BE,0} = 0$) by amount $0.5qV$ such that the total difference between the Fermi-energies is $E^{BE}_{f} - E^{TE}_{f} = qV$.
\begin{align}
    E^{TE}_{f} &= E^{TE,0}_{f} - 0.5qV \\
    E^{BE}_{f} &= E^{BE,0}_{f} + 0.5qV
\end{align}

We note that, in general, transitions will be field-dependent; a field-dependence introduces a time-dependence for the rates since the local electric field strength is dependent upon the spatial arrangement of charged defects -- which varies with time. For this reason, numerical modeling approaches require the solution to Poisson's equation at each time-step as has been done elsewhere~\cite{aldana_3D_2017,aldana_kinetic_2018,larcher_Microscopic_2012,zeumault_tcad_2021}. Towards compact-model development, as a simplification, everywhere the local field $\mathscr{E}$ is replaced with the average field $\bar{\mathscr{E}} = \frac{V}{t_{ox}}$. Consequently, the rates depend on the applied voltage but not on time.

\subsection{Electronic Conduction}

The current-density is computed from the following components: band-to-band tunneling between electrodes $J_t$, Ohmic conduction within the bands $J_{Ohmic}$ and bound-to-band tunneling from trapped electrons (i.e., trap assisted tunneling) $J_{tat}$. These are shown diagrammatically in Figure~\ref{fig:description}(b).
\begin{equation}
J = J_t + J_{Ohmic} + J_{tat} 
\end{equation}

The band-to-band tunneling component $J_t$ accounts for Fowler-Nordheim tunneling through the thin, insulating portion of the \ce{HfO2}~\cite{sze_physics_2006}.
\begin{equation}
    J_t = \frac{q^3}{8\pi h \Delta E^{TE/BE}}\mathscr{\bar{E}}^2\ex{-\frac{4\sqrt{2m^{ox}_e(\Delta E^{TE/BE})^3}}{3\hbar q\mathscr{\bar{E}}}}
\end{equation}
Alternatively, this can be refined with image-force correction using the general Simmons formula~\cite{simmons_Generalized_1963} as in~\cite{pickett_switching_2009}.

The Ohmic contribution is included based on charge neutrality considerations and considerations of information derived from prior literature. In our previous work, unoccupied oxygen vacancies were modeled as donors and occupied oxygen vacancies were modeled as acceptors~\cite{zeumault_tcad_2021}. The charge-state of the vacancy changed each capture/emission event, since dynamic updating of the charge state of the system for each discrete transition is possible using a KMC approach. Throughout these processes, charge neutrality can be satisfied, using the space charge of oxygen ions and an additional compensating positive charge to balance the additional negative charge created due to the capture of electrons injected from the electrodes.

To illustrate, we consider a process leading to electron capture by an oxygen vacancy ($*$ denotes a neutral site). Initially, Frenkel pair generation leads to neutral pairs of oxygen ions $O^{2-}$ and oxygen vacancies $V_O^{2+}$.
\begin{equation}
* * \to V_O^{2+} + O^{2-}
\end{equation}
After electron capture, negative oxygen ions remain, positive oxygen vacancies $N_{V_O^{2+}}$ are removed and negative oxygen vacancies $V_O^{2-}$ are created -- producing a net negative space charge equal to $-4e$ per capture event (assuming 4 electrons are captured~\cite{bradley_ElectronInjectionAssisted_2015}). 
\begin{equation}
V_O^{2+} + O^{2-} \to V_O^{2-} + O^{2-}
\end{equation}
Thus, by assuming that holes equal in concentration to $4N_{V_O^{2-}}$ are always present, bulk charge neutrality can be maintained within the oxide,
\begin{equation}
V_O^{2+} + O^{2-} \to V_O^{2-} + O^{2-} + 4h^+
\end{equation}
This assumes, that the oxygen ions $O^{2-}$ are present throughout all processes behaving as negative fixed charges. However, oxygen ions can be removed or introduced into the oxide via gettering processes at the electrode interfaces, establishing a spatial dependence, and our model does not incorporate the oxygen ion concentration as a state variable. Instead of introducing a new state variable, we assume that an electron concentration equal to $N_{V_O^{2+}}$ and a hole concentration equal to $N_{V_O^{2-}}$ are always present within the bands to ensure charge neutrality. We note that, since the oxygen vacancy state is a deep level state, the binding energy is stronger and the electron/hole states are not expected to behave as weakly-bound, fully delocalized states with their respective band mobility, especially at high defect concentrations. Instead, we assume a single effective mobility $\mu_{eff}$ parameter, with a drift current proportional to the total vacancy concentration.
\begin{equation}
    J_{Ohmic} = q\mu_{eff}\paren{N_{V_O^{2+}}+N_{V_O^{2-}}}  \mathscr{\bar{E}}
\end{equation}
Since this is a bulk component, it tends to become more important as the oxide thickness increases (reduced tunneling probability), as one might expect for highly defective insulators (e.g., \ce{SiN_x}); for thin films ($< \SI{5}{\nano\meter}$), the tunneling current dominates. Note, we model the Ohmic conduction as a simple phenomenological model, using a constant effective mobility parameter. If desired, this can be further modified to be temperature and field-dependent using a Poole-Frenkel mobility model as in disordered media~\cite{sze_physics_2006}. 
\begin{equation}
\mu_{eff} = \mu_0 \ex{-\frac{E_0}{kT}}\ex{\sqrt{\mathscr{E}}\paren{\frac{\beta}{T}-\gamma}}
\end{equation}
Where $\mu_0$, $E_0$, $\beta$, and $\gamma$ are fitting parameters~\cite{synopsys_TCAD_2019}. This is expected to be more significant for oxide memristors consisting of smaller bandgap oxides (e.g., \ce{TiO2}) presenting a smaller thermal barrier for local bound-to-band emission. However, it should be noted that, since Poole-Frenkel conduction is an additional emission component (to the bands within the oxide), the rate equation framework for the capture/emission process would need to be self-consistently modified if included, since all capture/emission processes effect the electron-occupancy of the oxygen vacancies. For simplicity, we do not consider such processes.

The trap-assisted-tunneling component is computed considering the current produced within a differential volume element centered about a discrete trap located within the oxide at $(x_t,y_t,z_t)$~\cite{schenk_advanced_1998}. In one dimension, we can write,
\begin{equation}
dJ_{tat} = q\mathscr{R}(y)dy \implies J_{tat} = q\int_{0}^{t_{ox}}\mathscr{R}(y)dy
\label{eqn:tat}
\end{equation}
where $\mathscr{R}$ is the net transition rate per unit volume, defined in terms of the electronic transition rates and density of states for traps $g(E_t)$, according to previous work~\cite{zeumault_tcad_2021}. 
\begin{equation}
    \mathscr{R}(y) = \int \paren{\frac{R_c^{BE}R_e^{TE} - R_c^{TE}R_e^{BE}}{R_c^{BE}+R_c^{TE}+ R_e^{BE}+R_e^{TE} }} g(E_t)dE_t
\end{equation}

%The current density can alternatively be expressed in terms of the spatial average of net transition rate $\bar{\mathscr{R}} \equiv \frac{1}{t_{ox}}\int \mathscr{R}(y)dy$
%%
%\begin{equation}
%J_{tat} = qt_{ox}\bar{\mathscr{R}}
%\end{equation}

Each electronic transition rate is position-dependent due to the tunneling factors. For a symmetric energy barrier at each electrode interface, the net transition rate will be maximum and sharply peaked for defects situated near the center of the oxide $y_t \approx 0.5t_{ox}$ (the centroid of a uniform defect profile), regardless of the defect distribution. If the integral in Equation~\ref{eqn:tat} can be approximated by its peak value, the spatial dependence can be eliminated. Replacing $y_t$ with $0.5t_{ox}$ yields the following,
\begin{equation}
    J_{tat} \approx qt_{ox}\int g(E_t) \paren{\frac{R_c^{BE}R_e^{TE} - R_c^{TE}R_e^{BE}}{R_c^{BE}+R_c^{TE}+ R_e^{BE}+R_e^{TE} }} dE_t
    \label{eqn:current-tat}
\end{equation}

This approximation effectively samples a narrow spatial range of the defect concentration (near the center of the oxide), regardless of the specific spatial distribution, since these will contribute more to the total integral in Equation~\ref{eqn:tat}.

\subsection{Self Heating}

Self-heating effects are approximated by assuming uniform Joule heating where the oxide volume dissipates heat via conduction to the external device temperature at $T_0$.
\begin{equation}
C_{device}\frac{dT}{dt} \approx I(t)V(t) - \frac{\kappa_{ox}A}{\Delta x}\paren{T(t)-T_0}
\label{eqn:selfheating}
\end{equation}
Starting with an initial temperature $T(0)=T_0$, equal to the device temperature, the temperature at the next time step is given in terms of the instantaneous temperature $T(t)$, thickness $t_{ox}$, specific heat capacity $c_{p,ox}$, mass density $\rho_{ox}$, thermal conductivity $\kappa_{ox}$ of the oxide, current density $J$, and ambient temperature $T_0$.
\begin{equation}
    T(t+\Delta t) = T(t) + \frac{\Delta t}{t_{ox}c_{p,ox}\rho_{ox}}\paren{J(t)V(t)-\frac{2\kappa_{ox}}{t_{ox}}\paren{T(t)-T_0}}
    \label{eqn:heat}
\end{equation}

Equation~\ref{eqn:heat} is the numerical solution to the time-dependent heat equation in one dimension, considering Joule heating and conductive heat losses to the electrodes (Equation~\ref{eqn:selfheating}). The factor $2$ is due to the assumption that heat dissipates from the center of the device over a distance equal to half the oxide thickness ($\Delta x = 0.5t_{ox}$). The model is approximate and, in practice, the quantity $\frac{2\kappa_{ox}}{t_{ox}}$ is the specific thermal conductance and is a fitting parameter. Use of self-heating creates internal temperature as an additional state variable for each memristor and, since temperature is not intrinsically bound, requires a window function to prevent temperatures from reaching absurd values. However, dynamic inclusion of self-heating may not always be necessary. For example, in oxide memristors having thin dielectrics, primary electronic conduction involves coupling traps within the dielectric to nearby electrodes via tunneling. Tunneling current, by contrast, produces a boundary condition for the current density, as opposed to a local current density within the oxide. Joule heating is therefore expected to occur near the electrodes – where tunneling electrons dissipate their excess kinetic energy into the contacts which behave as thermal reservoirs. Thus, for simplicity, if the electrode can be treated as an \textit{efficient} thermal reservoir, the dynamic effects of Joule heating might be approximated simply by observing how the model responds to static changes to the ambient temperature.

Close inspection of Equation~\ref{eqn:heat} shows that the ``device'' is taken as the oxide volume. Since the oxide volume constitutes a svery mall fraction of the thermal system (see SEM image within \cite{zhang_Experimental_2011}), a more realistic inclusion of self-heating considers that the oxide presents a negligible thermal mass in comparison to the electrodes and surrounding device structure which is much more efficient at conducting heat due to the lower thermal conductivity of the oxide. Without such a correction, inclusion of thermal models leads to absurdly high temperatures and/or nonsensical fitting parameters. For example, using the device geometry and thermal resistance ($R_{th} = \frac{\Delta_x}{\kappa A}$) reported in one comprehensive modeling approach of \ce{HfO2} memristors~\cite{guan_Switching_2012} the effective thermal conductivity of the \ce{HfO2} evaluates to \SI{100}{\watt\per\meter\per\kelvin}, which is $\approx$ 100 times larger than the measured value~\cite{panzer_Thermal_2009}. Equation~\ref{eqn:heat} is therefore modified to include a correction factor that accounts for the addenda heat capacity due to the bulkier elements in thermal contact with the oxide so that temperature effects can be modeled self-consistently using material parameters of the \ce{HfO2}.
\begin{equation}
    T(t+\Delta t) = T(t) + \frac{\Delta t}{t_{ox}c_{p,ox}\rho_{ox}CF}\paren{J(t)V(t)-\frac{2\kappa_{ox}}{t_{ox}}\paren{T(t)-T_0}}
    \label{eqn:heat-mod}
\end{equation}
Where $CF$ is a correction factor that takes into account the additional heat capacity of the system $C_{add}$.
\begin{equation}
C_{device} = C_{ox} + C_{add}
\end{equation}

It is simple to show (by adding additional thermal masses to the system defined by Equation~\ref{eqn:selfheating}) that the correction factor is defined as the ratio of the addenda heat capacity to the heat capacity of the oxide $C_{ox} = At_{ox}\rho_{ox}c_{p,ox}$ -- the addenda behaving as a heat sink; in this expression the electrode area is $A$ and the oxide volume is $At_{ox}$.
\begin{equation}
CF = 1 + \frac{C_{add}}{C_{ox}} \approx \frac{C_{add}}{At_{ox}\rho_{ox}c_{p,ox}} 
\end{equation}

Table~\ref{tbl:parameters} includes an estimate of the addenda heat capacity used in this work.

\subsection{Numerical Implementation}

Equations~\ref{eqn:master}, \ref{eqn:empty} and ~\ref{eqn:unoccupied} are solved iteratively by choosing an initial total vacancy concentration (Equation~\ref{eqn:total_vacancies}) and the system is updated at each time step. For convergence, the time step must be chosen to be smaller than the fastest transition rate. Since no rate can exceed $R_0$, the maximum timestep expected to converge is,
\begin{equation}
\Delta t \leq \frac{1}{R_0} = \SI{1e-13}{\second}
\end{equation}

This time step practically limits the total time duration of voltage segments. Since memristors can be programmed using nanosecond pulses, this is not particularly restrictive. We note that Equation~\ref{eqn:sum} allows one of the three state variables to be eliminated such that it is only necessary to track the occupied and unoccupied vacancy concentration with time (i.e., $N_{V_O^{2-}}$ and $N_{V_O^{2-}}$). The complexity of the compact model (and its computation speed) depends on the choice of density of states profile for the oxygen vacancy traps as will be shown. 

The implementation of the model is as follows:
\begin{itemize}
\item The equilibrium Fermi level is set to zero ($E^{TE,0}_{f} = 0$), and nominal band offsets and energy levels are defined with respect to the zero energy.
\begin{align}
V_{fb} &= \Phi_{TE} - \Phi_{BE}\\
E^{TE/BE,0}_c &= -\paren{\frac{3n^{TE/BE}_e\pi^2\hbar^3}{2\sqrt{2}\paren{m_e^{TE/BE}}^{3/2}}}^{2/3} \\
E_t &= \Phi^{BE} - \chi - E_{ion} - \frac{|V_{fb}|}{t_{ox}}y_t\\
\Delta E^{TE/BE} &= \Phi^{TE/BE} - \chi
\end{align}
\item A density of states profile is chosen to represent the oxygen vacancy distributions for all memristors being simulated. The energy levels of oxygen vacancies can be equivalently represented in terms of an ionization energy (relative to the conduction band) or the trap level (relative to the equilibrium Fermi level). The density of states profile is normalized such that the integral (over all energies) is equal to one. 
\begin{equation}
\int \hat{g}(E_t) dE_t = 1
\end{equation}
Therefore, the creation of additional defects scales all energy levels equally without modification to the shape of the density of states profile.
\item The initial concentration of oxygen vacancies is set. This value would represent the concentration of defects present in as-prepared oxide films. A nominal value used is \SI{1e10}{\per\centi\meter\cubed}.
\begin{align}
N_{V_O^{2+}}(0) &= N_{V_O^{2-}}(0) =\SI{1e10}{\per\centi\meter\cubed} \\
N_0(0) &= N_{sites} - N_{V_O^{2+}}(0) - N_{V_O^{2-}}(0)
\end{align}
\item The voltage versus time profile is defined. In this work, we consider positive and negative linear ramps applied in sequence to assess forming, reset, and set operations.
\item The Fermi level, Fermi functions, and transition rates are defined as functions of the instantaneous voltage and the trap level. These functions are called at each time step as the voltage changes.
\begin{align}
    E^{TE}_{f} &= E^{TE,0}_{f} - 0.5qV \\
    E^{BE}_{f} &= E^{BE,0}_{f} + 0.5qV \\
    E^{TE}_{c} &= E^{TE,0}_{c} - 0.5qV \\
    E^{BE}_{c} &= E^{BE,0}_{c} + 0.5qV \\
    f^{TE/BE} &= \frac{1}{\ex{\frac{E_t-E^{TE/BE}_f}{kT}}+1}
\end{align}
\item The state variables are updated using a forward Euler method to solve Equations~\ref{eqn:master}, \ref{eqn:empty} and \ref{eqn:unoccupied}.
\item The current components are computed at each time step from the state variables.
\item Calculated quantities are saved at specified intervals and written to a file.
\item A MATLAB script is used to implement device equations for testing/validation and Verilog-A code is implemented for circuit simulation in HSPICE, Cadence, and other simulation tools.
\end{itemize}

%According to Equation~\ref{eqn:master}, the use of any density of states profile other than a single energy level (delta profile) requires an integration at each time step, which is computationally inefficient, especially for circuits containing a large number of memristors. However, the use of a delta profile does not limit the ability of the model to fit experimental data, though it does limit the relevance of certain current components (as will be shown).

%The supplementary material contains additional figures comparing the simulation results with and without self-heating. 

%In what follows, we will show that the compact model provides quantitative agreement to experiment, has useful predictive capability of the effect of parameters on current-voltage characteristics, and can be used to simulate programming operations in 1T1R arrays.

\section{Results and Discussion}

\subsection{General Model Behavior}

\subsubsection{Density of States Profile}

In this section, we show how the total current density and current components vary with choice of density of states (DOS) profile for the oxygen vacancy states. Since oxygen vacancy states are electronically coupled to the electrodes, energy levels are necessarily broadened into continuous profiles. The ionization energy of a vacancy is also expected to vary with structural or chemical in-homogeneity, also leading to continuous profiles. For circuit simulation, the choice of DOS profile is important, since has an effect on the computational speed; any continuous DOS profile requires numerical integration at each time step for the calculation of capture/emission rates. The simplest case--the delta function DOS profile--is of particular interest for circuit simulation, since this model can be implemented without numerical integration at each timestep as will be shown. These considerations present an important tradeoff between model complexity and computational efficiency which can be selected by the needs of the circuit designer.

\begin{figure*}
\includegraphics[width=\textwidth]{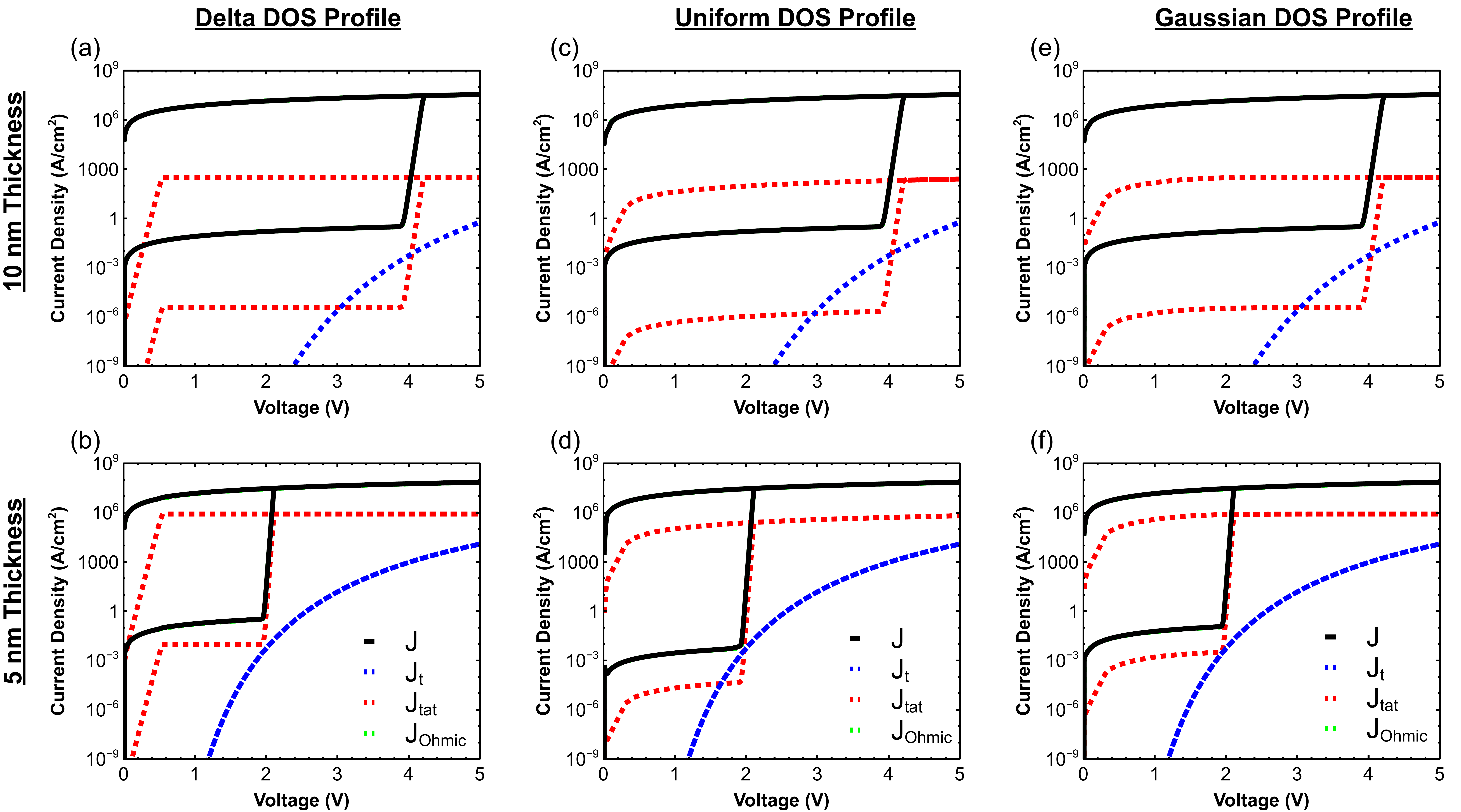}
\caption{Current density vs. voltage comparison for different density of states profiles and oxide thickness. Delta DOS profile (situated at \SI{2.9}{\electronvolt}) for an oxide thickness of (a) \SI{10}{\nano\meter} and (b) \SI{5}{\nano\meter}; Uniform DOS profile (throughout bandgap) for an oxide thickness of (c) \SI{10}{\nano\meter} and (d) \SI{5}{\nano\meter}; and Gaussian DOS profile (mean = \SI{2.9}{\electronvolt}, standard deviation = \SI{0.33}{\electronvolt}) for an oxide thickness of (e) \SI{10}{\nano\meter} and (f) \SI{5}{\nano\meter}. Nominal parameters for this simulation are as follows: $T_0=T(t)=\SI{300}{\kelvin}$, $t_{ox}=\SI{10}{\nano\meter}$, $N_{V_O^{2+}}(0)=N_{V_O^{2-}}(0)=\SI{5e11}{\per\centi\meter\cubed}$, $N_{sites}=\SI{4.38e19}{\per\centi\meter\cubed}$, $E_{a,gen} = \SI{7.35}{\electronvolt}$ (forming), $\sigma_0 = \SI{1e-14}{\centi\meter\squared}$.}
\label{fig:dos_comparison}
\end{figure*}

Figure~\ref{fig:dos_comparison} shows the current density versus voltage for the forming operation. Three different DOS profiles are compared, defined in terms of the ionization energies relative to the conduction band minimum $E_{ion}$: delta profile (situated at \SI{2.9}{\electronvolt}), uniform profile (throughout bandgap) and Gaussian profile (mean = \SI{2.9}{\electronvolt}, standard deviation = \SI{0.33}{\electronvolt}). Figures~\ref{fig:dos_comparison}(a), (c) and (e) correspond to an oxide thickness of \SI{10}{\nano\meter} whereas Figures~\ref{fig:dos_comparison}(b), (d) and (f) correspond to an oxide thickness of \SI{5}{\nano\meter}. 

For larger oxide thicknesses (i.e., \SI{10}{\nano\meter}), the Ohmic conduction model dominates the total current and is independent of the choice of DOS profile. This is expected, since tunneling current (band to band and trap-assisted) decreases exponentially with oxide thickness and bulk contributions become more dominant. Ohmic conduction, by contrast, is a local bulk component dependent upon the total defect concentration as opposed to the defect energy levels and/or their positions relative to the electrodes. 

As the oxide thickness reduces (i.e., \SI{5}{\nano\meter}), the importance of tunneling components increases, particularly during the transition from the pre-forming high-resistance state to post-forming low-resistance state. Furthermore, the voltage dependence of the trap-assisted tunneling current $J_{tat}$ becomes increasingly linear as the DOS profile is broadened from a delta profile to that of a uniform and gaussian profile. This is to be expected, since for small voltages, the current is proportional to the emission rate of a single trap state (if a delta profile is assumed), which is exponentially dependent on voltage. A continuous DOS allows for emission from multiple states at a given voltage, broadening the current-voltage dependence computed using Equation~\ref{eqn:current-tat}. In our previous work, a tunneling current was only considered, which produced linear current-voltage characteristics due to the varied contribution of defects in terms of their position and/or energy level~\cite{zeumault_tcad_2021}. We also note that the TAT current saturates for a single energy level (as seen in a Figures~\ref{fig:dos_comparison}(a), (b)) when the applied voltage is large enough that the thermal barrier to emission becomes zero (or negative). Beyond this voltage, the emission rate is equal to the pre-exponential factor.
\begin{equation}
R_{max} = n_e^{TE/BE}v^{TE/BE}_{th}\sigma_0\ex{-\frac{y_t}{y_0^{TE/BE}}}
\end{equation}

Importantly, these results suggest that, provided that the oxide thickness is not too small where tunneling significant occurs, the Ohmic conduction dominates the total current independent from the DOS profile. This allows a delta DOS profile to be used, which is more computationally efficient, despite being less representative of the physical reality as previously discussed. Importantly, setting the DOS profile as the delta profile,
\begin{equation}
g(E_t) = N_{V_O}\delta(E_{t}-E_{t0})
\end{equation} 
removes the integrals from Equations~\ref{eqn:master} and ~\ref{eqn:current-tat}.
\begin{equation}
\frac{dN_{V_O^{2-}}}{dt} = N_{V_O^{2+}}\paren{f^{BE}R_c^{BE}+f^{TE}R_c^{TE}}-N_{V_O^{2-}}\paren{(1-f^{BE})R_e^{BE}+(1-f^{TE})R_e^{TE}}
\label{eqn:master-delta}
\end{equation}

\begin{equation}
J_{tat} \approx qt_{ox}\paren{N_{V_O^{2-}} + N_{V_O^{2+}}} \paren{\frac{R_c^{BE}R_e^{TE} - R_c^{TE}R_e^{BE}}{R_c^{BE}+R_c^{TE}+ R_e^{BE}+R_e^{TE} }}
\end{equation}

\begin{figure*}
\includegraphics[width=\textwidth]{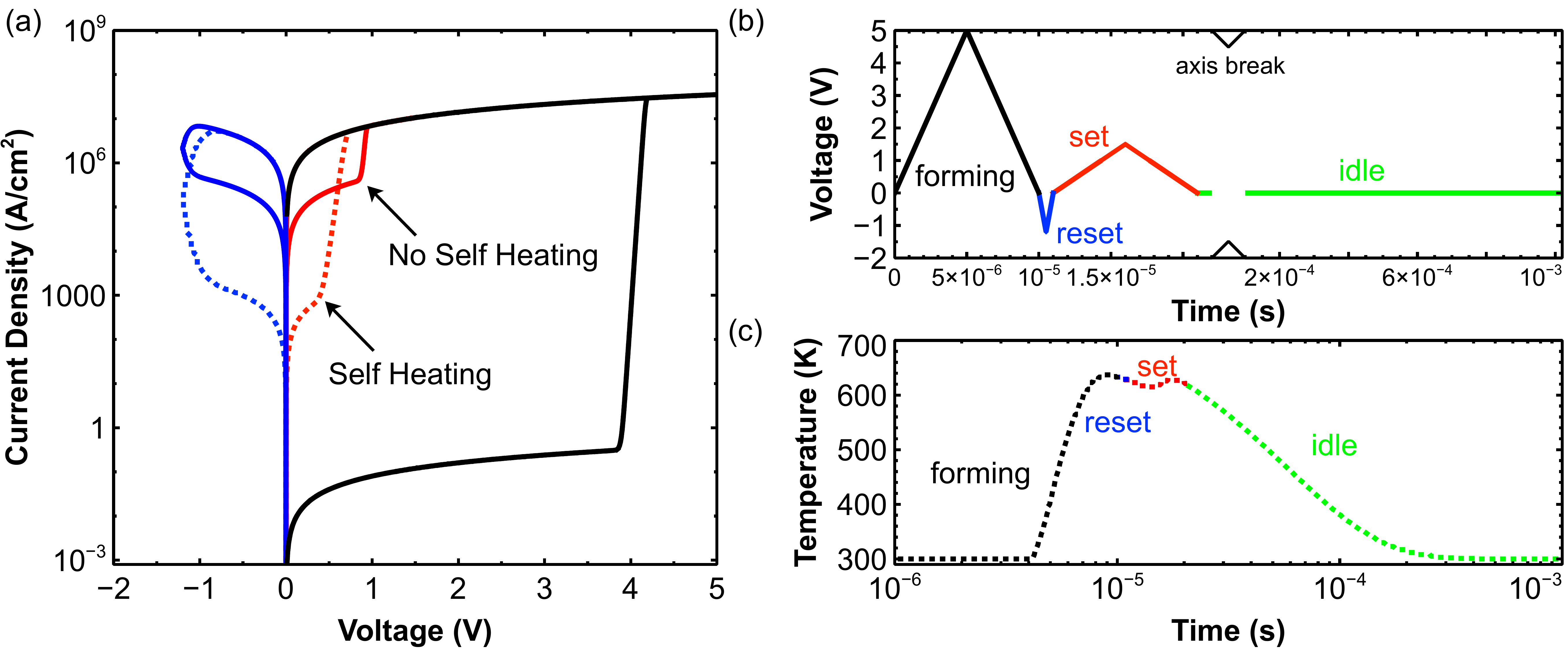}
\caption{(a) Nominal current-voltage characteristics computed by delta-shaped density of states profile, with (solid lines) and without self-heating (dashed lines) for an assumed \ce{TiN}/\ce{HfO2}/\ce{TiN} device structure. (b) Voltage versus time waveform. (c) Calculated temperature versus time profile due to self-heating. Parameters for this simulation are as follows: $T_0=T(0)=\SI{300}{\kelvin}$, $t_{ox}=\SI{10}{\nano\meter}$, $N_{V_O^{2+}}(0)=N_{V_O^{2-}}(0)=\SI{5e11}{\per\centi\meter\cubed}$, $N_{sites}=\SI{4.38e19}{\per\centi\meter\cubed}$, $E_{a,gen} = \SI{7.35}{\electronvolt}$ (forming), $E_{a,gen} = \SI{1.9}{\electronvolt}$ (set), $\sigma_0 = \SI{1e-14}{\centi\meter\squared}$.}
\label{fig:joule-heating}
\end{figure*}

Defining a net capture $R_c = f^{BE}R_c^{BE}+f^{TE}R_c^{TE}$ and emission rate $R_e = (1-f^{BE})R_e^{BE}+(1-f^{TE})R_e^{TE}$, the rate equations can be succinctly and intuitively expressed.
\begin{align*}
\frac{dN_{V_O^{2-}}}{dt} &= N_{V_O^{2+}}R_c-N_{V_O^{2-}}R_e \\
\frac{dN_{V_O^{2+}}}{dt} &= R_{gen}N_{0}-R_{rec}N_{V_O^{2+}}-N_{V_O^{2+}}R_c+N_{V_O^{2-}}R_e
\end{align*}
The expression for the empty states $N_0$ remains unchanged (Equation~\ref{eqn:empty}). Together, these represent a condensed, intuitive and more computatitonally efficient model that can represent the entirety of memristor characteristics as we show in Figure~\ref{fig:joule-heating}. Here, the model capability is compared with and without self-heating, showing forming, reset and set operations for a $\SI{10}{\nano\meter}$ oxide thickness. Self heating has the effect of increasing the extent to which the device resets to a high-resistance state. The higher temperature during set shifts the set voltage towards lower values (Figure~\ref{fig:joule-heating}(c)). We have assumed that the entire forming, reset and set operations occur in sequence without delay (Figure~\ref{fig:joule-heating}(b)). If there is time between programming cycles (above $\approx$ \SI{1}{\milli\second}), the device will cool back to the ambient temperature $T_0$ (Figure~\ref{fig:joule-heating}(c)) during ``idling'' and the cumulative-effects of self-heating would be diminished. The time constant for recovery to the ambient temperature can be calculated from setting the Joule-heating term in in Equation~\ref{eqn:selfheating} to zero, obtaining the familiar Newton's cooling law result.
\begin{equation}
T(t) = T_0 + T(t_0)\ex{-\frac{t-t_0}{\tau}}; \qquad \tau = \frac{t_{ox}C_{device}}{2\kappa_{ox}A}
\end{equation}
Using the parameter values in Table~\ref{tbl:parameters}, $\tau \approx \SI{60}{\micro\second}$. Assuming that the temperature has recovered within a total time equal to $3\tau \approx \SI{0.18}{\milli\second}$, the results shown in Figure~\ref{fig:joule-heating}(c) are sensible. The exact rate of recovery is expected to differ due to differences in memristor and electrode materials and device geometry.

In the following sections, we apply this delta-profile DOS model to gain an intuitive understanding of state variable dynamics, parameter dependencies and predictability, as well as making comparisons to experimental data.

\subsubsection{State Variable Dynamics}

In this section, we show how the state variables change with time. The essential aspects of the memristor model will be shown using the forming operation, to avoid restating similar trends that occurs from simply changing the polarity/magnitude of voltage. Importantly, during forming, the current exhibits a sharp increase from a low value to a high value at a certain voltage and the current-voltage characteristics exhibit hysteresis and a zero-crossing. In our model, an increase in current stems from an increase in oxygen vacancy concentration--through defect generation and electron capture. By contrast, a decrease in current stems from a decrease in vacancies--through recombination of unoccupied vacancies and electron emission from occupied vacancies. 

\begin{figure*}
\includegraphics[width=\textwidth]{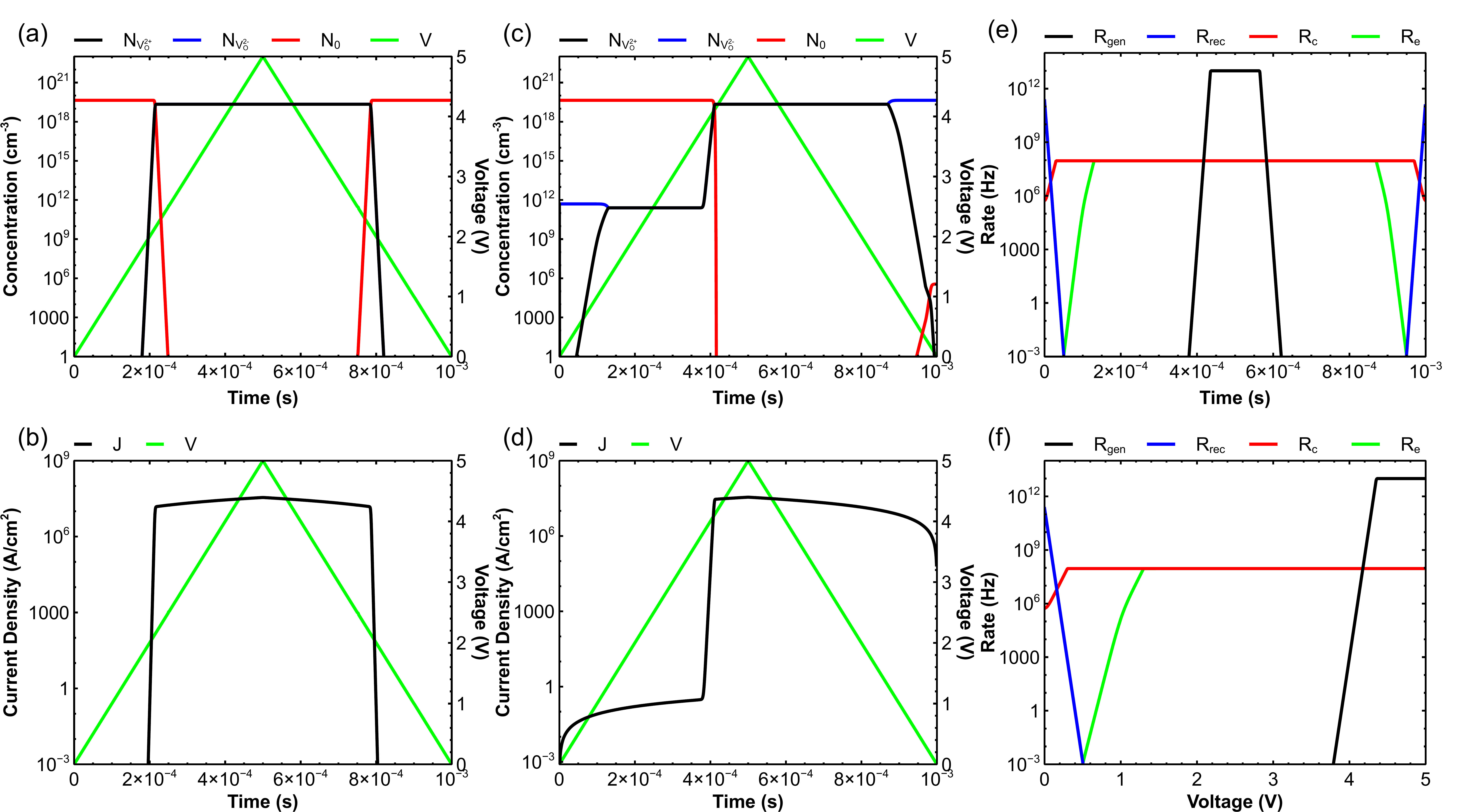}
\caption{State variables and current density versus time for DC (a),(b) and transient (c),(d) simulations. Transition rates versus time (e) and voltage (f). Note: $R_c$ and $R_e$ are the total capture and emission rates for both electrodes. Nominal parameters for this simulation are as follows: $T_0=T(t)=\SI{300}{\kelvin}$, $t_{ox}=\SI{10}{\nano\meter}$, $N_{V_O^{2+}}(0)=N_{V_O^{2-}}(0)=\SI{5e11}{\per\centi\meter\cubed}$, $N_{sites}=\SI{4.38e19}{\per\centi\meter\cubed}$, $E_{a,gen} = \SI{7.35}{\electronvolt}$ (forming), $\sigma_0 = \SI{1e-14}{\centi\meter\squared}$.}
\label{fig:dc_transient_comparison}
\end{figure*}

Figure~\ref{fig:dc_transient_comparison} shows the evolution of the state variables with time for DC and transient conditions for a forming operation from \SI{0}{} to \SI{5}{\volt}. As shown in Figure~\ref{fig:dc_transient_comparison}(a),(b), the theoretical DC solution does not show hysteresis; the absence of hysteresis is evidenced by the fact that the state variables and current density are symmetric with respect to the applied voltage versus time waveform. This observation is common among physics-based memristor models, which has led to debates regarding whether these devices can, strictly speaking, be classified as memristors. However, as argued by Wang et al~\cite{wang_well-posed_2016}, since the time scales needed to reach the theoretical DC solution are astronomically large, such arguments are not restrictive in practice and may be only of academic interest. 

Furthermore, the lack of DC hysteresis suggests the low-resistance state of a memristor is metastable; the high-resistance state can be recovered by simply raising the device temperature. The DC current is negligible for small voltages because vacancies are unstable to recombination and/or electron emission in the infinite time (i.e., DC) or infinite temperature limit. This has been shown experimentally by Kumar et al~\cite{kumar_Conduction_2016}--oxygen diffusion ultimately leading to recombination of Frenkel pairs at vacancy sites with full reset occuring over a time scale of minutes at temperatures of $\approx$ \SI{550}{\kelvin}. In their work, the thermal barrier to recovery of the pre-forming high resistance state is \SI{0.7}{\electronvolt}, which is comparable to the thermal barrier to oxygen ion diffusion~\cite{zeumault_tcad_2021}. Assuming an attempt frequency of \SI{10}{\tera\hertz}, at a temperature of \SI{550}{\kelvin} the time scale for diffusion would be $\approx$ \SI{0.2}{\micro\second} which is much faster than the time scale observed for reset. Our model also considers electron emission as a parallel process whereby a conductive filament becomes more susceptible to dissolution (see Equation~\ref{eqn:master}), since it has been shown that electron-occupied vacancies are more stable to recombination~\cite{bradley_ElectronInjectionAssisted_2015}. Using nominal parameters, our model calculates an emission rate of $\approx$ \SI{1}{\milli\hertz} to an assumed \ce{TiN} electrode from a trap with ionization energy of \SI{3}{\electronvolt} evaluated at low positive voltages $\approx$ \SI{0.1}{\volt}. The time scale of this process is of the order of minutes and is more consistent with the experimental data in~\cite{kumar_Conduction_2016} than a recombination process acting alone. It would be interesting to repeat this experiment for different electrode work functions to see if the thermal barrier is work function dependent and, therefore, if an electron-emission process has a rate-limiting effect on the recovery process.

Transient simulations, by contrast, do show hysteresis. Ramped conditions favor electron capture as opposed to emission as a precursor for recombination, which is a comparatively much slower process for deep traps; electron capture forms a negatively charged vacancy which is more stable to recombination~\cite{bradley_ElectronInjectionAssisted_2015}. This is shown in Figure~\ref{fig:dc_transient_comparison} (c), (d) for a ramp rate of \SI{10}{\volt\per\milli\second}. This ramp rate is much slower, by many orders of magnitude, than ramp rates used during programming cycles in practice (e.g., write operations) which is of the order of $>\SI{1}{\volt\per\nano\second}$. As shown in Figure~\ref{fig:dc_transient_comparison}(f), the emission rate remains lower than the capture rate (by many orders of magnitude) at low voltages until $\approx$ \SI{1.3}{\volt}. At this voltage, the Fermi level of the top electrode is approximately aligned with the trap level at $E_{ion}\approx \SI{3}{\electronvolt}$, increasing the emission rate to become comparable to the capture rate. When the capture and emission rates are both high--for positive voltages--the rate of recombination is low, preventing reset; for negative voltages the recombination rate is high, leading to reset. Eventually, the generation rate becomes substantial at large positive voltages, causing the vacancy concentration to increase. On the reverse sweep, the emission rate falls off, stabilizing a large concentration of occupied vacancies and a corresponding low resistance state. Thus, for all practical purposes, the model can be regarded as capable of reproducing hysteresis, and would accurately predict the vanishing of hysteresis at temperatures exceeding $\approx$ \SI{550}{\kelvin} and for pulses exceeding minutes in duration.  

In both cases (DC and transient), the transition from a pre-forming high resistance state to a post-forming low resistance state is coincident with a reduction in the concentration of empty states $N_0$ and corresponding increase in vacancies $N_{V_O^{2+}}$ and $N_{V_O^{2-}}$. Qualitatively, this is similar in effect to a reduction in tunneling gap according to previous models~\cite{guan_SPICE_2012}. On the forward sweep, due to the large difference in the rate of electronic processes and atomic processes, the increase in $N_{V_O^{2-}}$ tracks the exponential increase in $N_{V_O^{2+}}$ since electron capture is energetically favorable from the bottom electrode for positive applied voltage $V=V_{TE}-V_{BE} > 0$. On the reverse sweep, the concentration of occupied vacancies is stable due to a sizeable thermal barrier to emission; the barrier is stable since the trap level is lower than the Fermi energy of either a electrode (i.e., $E_t - E_f < 0$). The new value of $N_{V_O^{2+}}$ reached once the voltage at the top electrode returns to zero is based on a balance between the rate of recombination of Frenkel pairs $R_{rec}$ and the combined rate of electron capture $R_c$ according to Equation~\ref{eqn:unoccupied}. This can be seen from the increase in the concentration of empty states and a decrease in unoccupied vacancies towards the end of the pulse $\SI{0.9}{\milli\second}< t < \SI{1}{\milli\second}$.

\begin{figure*}
\includegraphics[width=\textwidth]{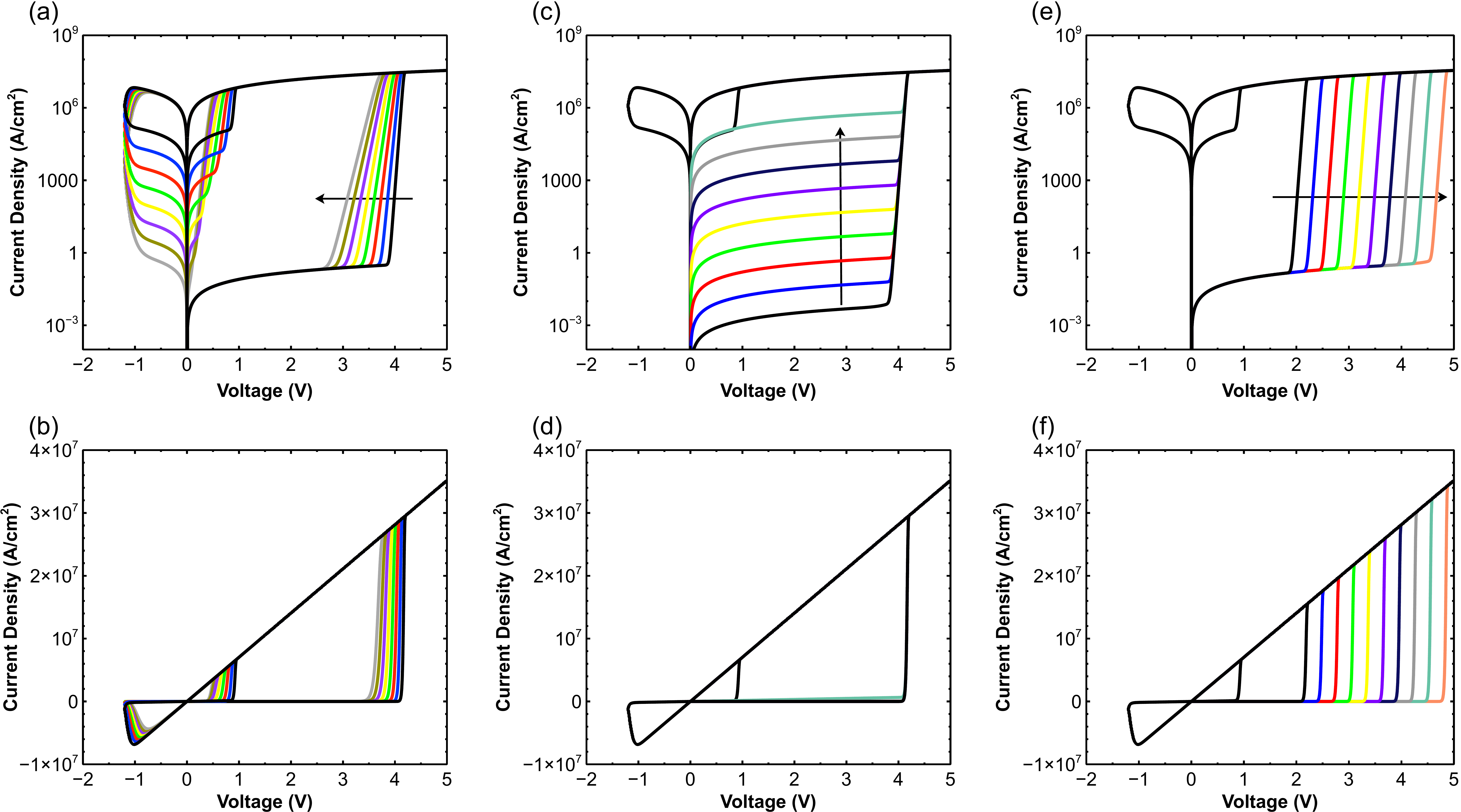}
\caption{(a),(b) Effect of temperature: temperature was varied from \SI{300}{\kelvin} to \SI{1000}{\kelvin} in \SI{100}{\kelvin} increments.  (c),(d) Effect of initial initial vacancy concentration: initial vacancy concentration was varied logarithmically from \SI{1e10}{\per\centi\meter\cubed} to \SI{1e18}{\per\centi\meter\cubed} in 1 decade increments. (e),(f) Effect of activation enthalpy of Frenkel pair generation:  enthalpy was varied from \SI{4}{\electronvolt} to \SI{8.5}{\electronvolt} in \SI{0.5}{\electronvolt} increments. The current density is shown in both logarithmic (top) and linear (bottom) scale.}
\label{fig:parameter_comparison}
\end{figure*}

We note that a particular strength of our approach is that, based on Equation~\ref{eqn:sum}, the state variables are bound between the atomic density $N_{sites}$ and zero, provided the rates are finite. As pointed out by Wang et al~\cite{wang_well-posed_2016}, the use of a physical description based on a tunneling gap as a state variable requires the use of a window function to prevent the gap from diminishing below zero or extending beyond the oxide thickness. Using the oxygen vacancy concentrations as state variables eliminates the need for such a window function.

\subsubsection{Predictive Capability of Model}

Figure~\ref{fig:parameter_comparison} shows the predictive capability of the model for variable temperature (a),(b), intitial vacancy concentration (c),(d) and activation enthalpy (e),(f). Pulse durations used were \SI{10}{\micro\second}, \SI{1}{\micro\second} and \SI{10}{\micro\second} for forming, reset and set respectively. 

Temperature was varied from \SI{300}{\kelvin} to \SI{1000}{\kelvin}. Temperature changes affect the onset forming and set voltage and the resistance level following reset. At higher temperatures, the voltage required for forming is reduced; this indicates an increase in the rate of forming kinetics as predicted by the thermochemical model of dielectric breakdown~\cite{mcpherson_thermochemical_2003}; this has also been confirmed experimentally, by measuring the time required to form at various temperatures~\cite{lorenzi_forming_2013}. Importantly, a higher temperature increases the rate of electron emission from occupied vacancies, creating more unoccupied vacancies which are available for recombination with oxygen ions, therefore increasing the rate of reset.

The initial vacancy concentration was varied from \SI{1e10}{\per\centi\meter\cubed} to \SI{1e18}{\per\centi\meter\cubed}. An increase in vacancy concentration has the effect of modifying the high-resistance state of the oxide film prior to forming only; as expected, a higher defect concentration leads to a higher current density and a lower resistance level; this parameter can be adjusted to reflect differences in preparation techniques used to deposit oxide thin films which will change due to processing parameters. 

Lastly, the activation enthalpy of Frenkel pair generation is modified from \SI{4}{\electronvolt} to \SI{8.5}{\electronvolt}. As expected from Equation~\ref{eqn:thermochemical}, the voltage required for forming linearly increases with an increase in activation enthalpy. Again, this follows from the thermochemical model of dielectric breakdown~\cite{mcpherson_underlying_1998,mcpherson_thermochemical_2003}; this parameter can be adjusted to match forming voltages extracted from experimental data sets. To match set voltages, a separate activation enthalpy is used (not shown). 

In the next section, we illustrate choice of these parameters using an experimental dataset.

% Compact Model Validation

\begin{figure*}
\includegraphics[width=\textwidth]{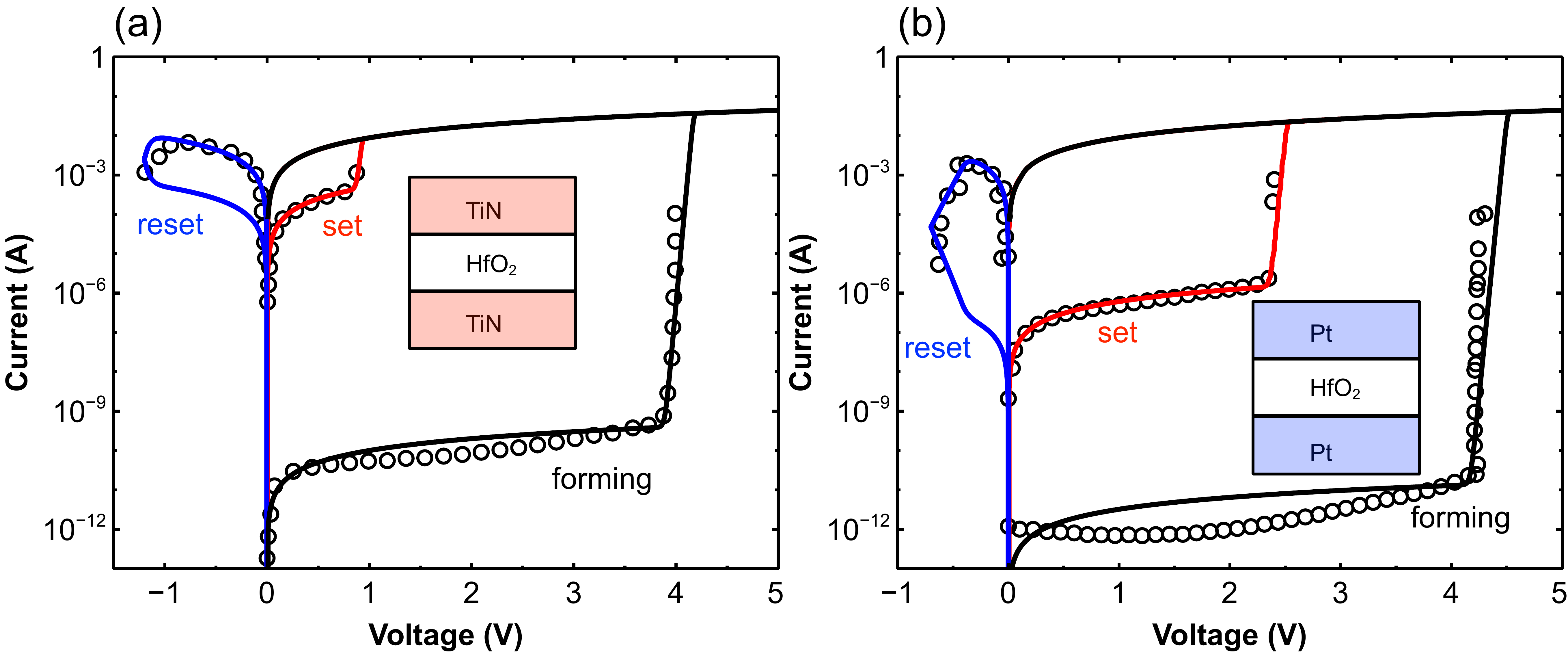}
\caption{
Comparison of current-voltage characteristics from compact model and experimental data for two memristor devices with identical geometry (area = \SI{1.25e-13}{\meter\squared}, oxide thickness = \SI{10}{\nano\meter}) having different electrode structures, with identically-processed \ce{HfO2} layer~\cite{lorenzi_forming_2013}: (a) \ce{TiN}/\ce{HfO2}/\ce{TiN} and (b) \ce{Pt}/\ce{HfO2}/\ce{Pt}. The compact model (solid lines) shows a reasonable match with the experimentally observed characteristics (open circles).
%Modeled Current-voltage characteristics of (a) a \ce{TiN}/\ce{HfO2}/\ce{TiN} memristor \cite{lorenzi_forming_2013} and (b) a \ce{TiN}/Ti/\ce{HfO2}/\ce{TiN} memristor \cite{Aziza2019}. Our compact model (solid lines) shows a reasonable matching with the experimentally observed characteristics. (c) A list of the parameters used in our model to match with these experiments. To match with the experiment I, total simulation time of \SI{200} {\micro\second} and ramp rate of \SI{0.07} {\volt/\micro\second} are used and for experiment II, these values are \SI{125} {\micro\second} and \SI{0.12} {\volt/\micro\second}, respectively.
}
\label{fig:experimental_comparison}
\end{figure*}

\subsection{Model Validation and Testing}

\subsubsection{Comparison to Experimental Data}

Having demonstrated the general behavior of the model, we now utilize this framework to develop a circuit-compatible compact model in Verilog-A based upon a delta DOS profile. 

First, we compare our compact model (using delta DOS profile) with two different \ce{HfO2}-based memristors having different electrode materials in which the \ce{HfO2} layer was identically processed using atomic layer deposition at \SI{350}{\celsius}~\cite{lorenzi_forming_2013}. This dataset was selected based on the availability of processing and characterization data which provides details regarding the structure of the \ce{HfO2} layer in addition to the electrode interfaces~\cite{cabout_Role_2013,jorel_Physicochemical_2009a,lorenzi_forming_2013}. The memristors have identical geometry (area = \SI{1.25e-13}{\meter\squared}, oxide thickness = \SI{10}{\nano\meter}) with an electrode thickness of \SI{25}{\nano\meter}. The devices have symmetric electrode configurations \ce{TiN}/\ce{HfO2}/\ce{TiN} (shown in Figure~\ref{fig:experimental_comparison}(a)) and \ce{Pt}/\ce{HfO2}/\ce{Pt} (shown in Figure~\ref{fig:experimental_comparison}(b)). Apart from an apparent negative differential resistance in the as-prepared (pre-forming) state present in both but particularly pronounced in the \ce{Pt}/\ce{HfO2}/\ce{Pt} device (likely due to initial charging of interface traps~\cite{rofan_Stressinduced_1991,chen_Reliability_}), the devices are well-behaved. We note that other devices prepared using the same process show similar forming voltages and reduced current magnitudes without NDR behavior~\cite{cagli_Experimental_2011} and are therefore within the large statistical spread of these devices~\cite{lorenzi_forming_2013} and qualitatively identical. Electrode parameters were used from prior literature reports for \ce{TiN}~\cite{lima_Titanium_2012,solovan_Electrical_2014,walker_estimate_1998} and \ce{Pt}~\cite{schaeffer_Contributions_2004,fischer_Mean_1980,hoffmann_Electrical_1976} thin-films.

The simulated current-voltage characteristics for these two \ce{HfO2} memristors are compared with experimental data reported in~\cite{lorenzi_forming_2013} in Figure~\ref{fig:experimental_comparison} (a) and (b). As evidenced by the smaller current in the pre-forming state, the \ce{Pt} electrode device is fit using a smaller initial oxygen vacancy concentration. The authors report a higher oxygen vacancy concentration in \ce{HfO2} films fabricated on \ce{TiN} (and \ce{Ti}) as opposed to films grown on \ce{Pt}, attributed to the formation of a titania suboxide (\ce{TiO_x}) interfacial layer. This is sensible from a thermodynamics perspective given the large formation enthalpy of \ce{TiO2} and the associated gettering behavior of \ce{Ti}~\cite{stout_Gettering_1955}. Importantly, from a modeling perspective, this justifies the increase in the initial oxygen vacancy concentration in \ce{HfO2} films deposited on \ce{TiN}. 

The forming and set voltages are higher for the \ce{Pt} electrode device than the \ce{TiN} electrode device. The authors attribute the reduction in forming voltage to a reduced formation enthalpy of Frenkel pairs within the vacancy-rich interfacial layer adjacent to the \ce{TiN} electrodes. The lower formation enthalpy may reflect the fact that nearest-neighbor pairs and clusters of oxygen vacancies are more stable than point defects~\cite{bradley_ElectronInjectionAssisted_2015}; vacancy clusters ultimately produce what is regarded as a ``conductive filament" and are present after initial forming. For the forming data shown in Figure~\ref{fig:experimental_comparison}(a), the forming voltage is on the higher end of the statistical variation reported for \ce{Ti}/\ce{TiN} based electrodes, which can be much lower by a couple volts~\cite{cabout_Role_2013}. Although this explanation is consistent with theoretical investigations in the formation of vacancy clusters, it should be noted that the formation of positive space charge (e.g., unoccupied oxygen vacancies) near the anode/TE or negative space charge (e.g., negative oxygen vacancies) near the cathode/BE  would be expected to lead to a local field enhancement within the oxide--both reducing the forming voltage. Additionally, the stoichiometric changes (when corrected for improvements to polarizability due to crystallization~\cite{venkataiah_Oxygen_2021}) can possibly lead to changes in forming and set voltages according to the thermochemical model (Equation~\ref{eqn:thermochemical}). 
\begin{equation}
\mathscr{E}_{bd} = \frac{E_{a,gen}}{|\vec{p}|\paren{\frac{2+\epsilon_r}{3}}}
\end{equation}
Thus, either a reduction in formation enthalpy or an increase in polarization (e.g., relative permittivity or molecular dipole moment) could be used to describe the differences in transition voltages in these datasets. Here, to describe the forming/set operations, the activation energy for Frenkel pair generation was reduced. We note that the transition voltages are sweep rate dependent; we have used linear voltage ramps over the indicated voltage ranges with a total time interval per segment of \SI{10}{\micro\second} for forming and set and \SI{1}{\micro\second} for reset operations in both devices corresponding to ramp rates between \SI{0.1}{\mega\volt\per\second} and \SI{1.2}{\mega\volt\per\second}, consistent with the reports for this dataset~\cite{cabout_Role_2013}.

The authors also report the presence of different crystalline phases between the two \ce{HfO2} films -- films grown on \ce{Ti} or \ce{TiN} are predominantly orthorhombic whereas films grown on \ce{Pt} are predominantly monoclinic. The monoclinic phase is the least polarizable among the polymorphs of \ce{HfO2} with a relative permmitivity of 18 and 29 for the cubic phase~\cite{zhao_Firstprinciples_2002}. The dominant orthorhombic phase observed on \ce{TiN} is therefore consistent with a reduced forming/set voltage relative to monoclinic films grown on \ce{Pt} without need for reducing the activation enthalpy. The oxygen vacancy formation energy in both phases is similar at the same Fermi level position~\cite{wei_Intrinsic_2021}, so the ionization energy is kept the same when fitting the datasets. Only a slight increase in capture cross section (from \SI{1e-14}{\centi\meter\squared} to \SI{2e-14}{\centi\meter\squared}) was needed to fit the \ce{Pt} devices. 

In sum, for both datasets, our compact model demonstrates a close match with the experiments in all operating conditions (forming, set, and reset) of a memristor. The many physical considerations emphasize the need for detailed characterization in order to leverage the model's predictive capability and justify changes to particular parameters over others. Table~\ref{tbl:experimental-fit} lists the parameter values adjusted to achieve good agreement between the modeled and measured data.

\begin{table}
\begin{center}
\begin{tabular}{ccccccc}
\hline
 & $N_{V_O^{2+}}(0)$[\SI{}{\per\centi\meter\cubed}] & $N_{sites}$[\SI{}{\per\centi\meter\cubed}] & $E_{ion}$[\SI{}{\electronvolt}] & $\sigma_0$[\SI{}{\centi\meter\squared}] & $E_{a,gen}^{forming}$[\SI{}{\electronvolt}] & $E_{a,gen}^{set}$[\SI{}{\electronvolt}]\\
\hline
\ce{TiN}/\ce{HfO2}/\ce{TiN} & \SI{5e11}{} & \SI{4.38e19}{} & \SI{2.957}{} & \SI{1e-14}{} & \SI{7.35}{} & \SI{1.90}{}\\
\ce{Pt}/\ce{HfO2}/\ce{Pt} & \SI{1.61e10}{} & \SI{4.38e19}{} & \SI{2.957}{} & \SI{1.7e-14}{} & \SI{7.91}{} & \SI{4.56}{}\\
\hline
\end{tabular}
\end{center}
\caption{Summary of parameter used in fitting the data in Figure~\ref{fig:experimental_comparison}.}
\label{tbl:experimental-fit}
\end{table} 

\subsubsection{Simulation of 4x4 1T1R Nonvolatile Memory Array}

%Array Simulation

\begin{figure*}
\includegraphics[width=\textwidth]{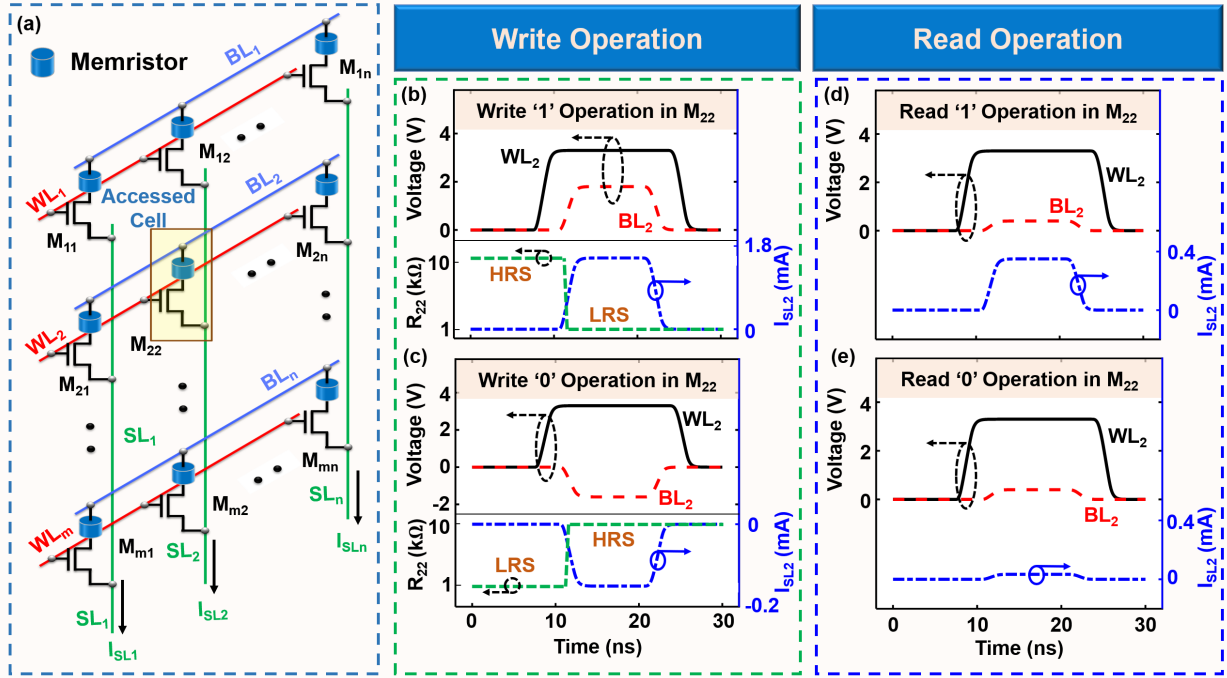}
\caption{(a) Schematic of a 1T1R array where we perform the write and read operations in the (2, 2) cell. Top Panels show the time dynamics of WL and BL biases for (a) write '1', (b) write '0', (c) read '1', and (d) read '0' operations in the accessed cell. Bottom panels show the corresponding effects of the WL and BL biases on the accessed cell during a) write '1', (b) write '0', (c) read '1', and (d) read '0' operations.}
\label{fig:array}
\end{figure*}

Finally, to demonstrate the usability of our model and prove that our model can be self-consistently coupled with other circuit elements, we simulate a $\SI{4} {\times} {4}$ 1-transistor 1-memristor (1T1R) array (shown in Fig.~\ref{fig:array}(a)). In our simulation, we use the \SI{65}{\nano\meter} DGXFET models available in the IBM CMOS 10LPe process for the access transistors and our compact model calibrated with the experiment \cite{Aziza2019} for the memristor-based memory elements. In memristor-based memory systems, two resistance levels -- HRS (high resistance state) and LRS(low resistance state) -- are used to define the memory states (logic '0' and logic '1', respectively). Here, we demonstrate the write and read operations in the (2, 2) cell of the array. To access the cell, we fist apply the suitable voltage to the corresponding WL $(WL_2)$ to turn the access transistor ON (see top panels of Figs.~\ref{fig:array}(b), (c), (d), and (e)). Then, to write into (read from) the accessed cell, we apply the suitable write voltage (read voltage) to the BL $(BL_2)$. The remaining WLs and BLs are kept at ground. 

Figures ~\ref{fig:array}(b) and (c) show that with the suitable biases at WLs and BLs, the resistance of the accessed cell switches from one state to another: HRS $\rightarrow$ LRS during write '1' and LRS $\rightarrow$ HRS 
during write '0' operations. During the read operation, for a same amount of voltage at the BL, we observe two levels of current based on the stored memory state in the cell--high current for logic '1' and low current for logic '0' states. A simple current sense amplifier \cite{Chang2013,aziz_Low_2017} with a suitable reference can be used for the sensing purpose thanks to the distinct current levels corresponding to two memory states.

\section{Conclusions}

In summary, a new compact model for oxide memristors was presented, based on the use of oxygen vacancy concentration as state variable. The theoretical model is based upon a recent manuscript which combined rates of atomic processes (e.g., Frenkel pair generation and recombination, diffusion of point defects) with those of electronic processes (electron capture and emission), in a kinetic Monte Carlo simulation approach. The compact model was validated using bulk parameters for \ce{HfO2}, though is of general utility to a wide range of single phase, insulating oxides or polymorphs which can be described using ``effective'' bulk-averaged parameters. The dynamic evolution of the state variables -- the concentrations of occupied and unoccupied oxygen vacancies -- was shown to be capable, both qualitatively and quantitatively, of reproducing the essential switching characteristics of oxide memristors. In particular, the increase (decrease) in oxygen vacancy concentration is qualitatively similar in effect to the more familiar reduction (expansion) of the tunnel gap which has been used in existing compact models.  Key strengths of this approach include: the state variables are bound between zero and the atomic density of the oxide without the use of a window function -- needed for preventing tunnel gaps from becoming negative or exceeding the oxide thickness; the model makes few assumptions regarding the ``filament" geometry, being chiefly based upon the number density of oxygen vacancy defects as opposed to their geometry; the model provides an intuitive description of resistive switching that is consistent with retention improvements observed in lower work-function metal electrodes (e.g., \ce{Ti}/\ce{TiN}) compared to \ce{Pt} due to an increased emission barrier which stabilizes the low-resistance state; the model can be easily refined via inclusion of image force correction or Poole-Frenkel modeling of high-resistance states without loss of generality.

As a test of its practical utility, the compact model was implemented in Verilog-A, verified using well-defined experimental data-sets with different electrode geometries, and tested in circuit simulation using a 4x4 one-transistor, one-resistor (1T1R) memory cells, representative of a small, nonvolatile memory array. We anticipate that this model will serve as an alternate description of memristor switching in terms of the concentration of defects as opposed to the state of the filament. Such a description will be particularly useful as new and ongoing experimental data emerges to suggest new or confirm existing geometrical descriptions of conductive filament shapes.

\section*{Data Availability Statement}

The data that support the findings of this study are available from the corresponding author upon reasonable request.

\bibliography{references-az,references,references-shamiul}

\end{document}